%% file: coul-ren-prl.tex
\documentclass[prl,aps,twocolumn,superscriptaddress,floatfix]{revtex4-1} 
\usepackage{epsf}
\usepackage{epsfig}
\usepackage{graphics}
\usepackage{amssymb}
\usepackage{hyperref}
\usepackage{color}
\begin{document}

\title{Reduction of electron repulsion and enhancement of $T_{c}$ in small diffusive superconducting grains}

\author{Sabyasachi Tarat}
\affiliation{Physics Department, Ben Gurion University, Beersheba 84105, Israel}
\affiliation{Department of Condensed Matter Physics, Weizmann Institute of Science, Rehovot 76100, Israel}

\author{Yuval Oreg}
\affiliation{Department of Condensed Matter Physics, Weizmann Institute of Science, Rehovot 76100, Israel}

\author{Yoseph Imry}
\affiliation{Department of Condensed Matter Physics, Weizmann Institute of Science, Rehovot 76100, Israel}

\date{\today}

\begin{abstract}
The superconducting properties of small metallic grains has been a topic of active research for half a century now.
Early experiments demonstrated a remarkable rise in the critical temperature, $T_{c}$, with reducing grain size in a variety of materials. 
 In two dimensional diffusive superconductors, $T_{c}$ is decreased due to enhanced Coulomb repulsion.
We propose that in finite size grains, the diffusive enhancement  of the Coulomb repulsion is weakened 
and leads ultimately to an increase in $T_{c}$  in isolated, disordered two dimensional grains.
Our mechanism is superimposed on the possible enhancement in $T_{c}$ due to the change in the density of states of finite size systems.
\end{abstract}

\maketitle
{\it Introduction}: The superconducting properties of materials composed of small metallic grains has been a topic of 
enduring interest for more than five decades now, beginning with the pioneering theoretical work of Anderson in 1959 \cite{and-tr}. 
In the late 1960s, a series of experiments on thin films of granular $Al$, $Sn$, $In$ etc.\cite{cohen12,deutscher1} 
found a remarkable enhancement in their transition temperatures as the grain size was reduced
~\cite{fn1,bh-paps,burmistrov}.
Later, improvements in electron tunneling methods enabled measurements on single grains \cite{brt12},  and
showed that the enhanced $T_{c}$ in these grains was accompanied by an enhanced single particle gap 
compared to the bulk value. Recent experiments \cite{deutscher2} in dense grain arrays seem to be consistent with 
these older observations. 

While initial explanations of this increase included proposals 
such as a surface enhancement of electron-phonon interactions \cite{cohen12}, later theories  
have tried to explain this in terms of various finite size effects \cite{fin-paps}, that become important 
with reducing grain size, as the single particle level spacing, $\delta$, increases. 
In relatively clean systems, this could lead to an enhancement
in the density of states (DOS) at the Fermi level, resulting in an increasing $T_{c}$ with reducing size, until the grain
becomes small enough such that $\delta \sim \Delta$, where $\Delta$ is the 
superconducting gap. Below this minimum size, a coherent superconducting state can no longer
be formed in a single grain and the $T_{c}$ disappears \cite{and-tr, tc-vanish-paps}.
In dirty or irregularly shaped grains, on the other hand,
the interplay of disorder, electron-electron repulsion and finite size effects brings non-trivial physics into play,
as we explain in detail below.

In a conventional superconductor, the attractive interaction responsible for 
superconductivity is mediated by electron phonon interactions. When an electron
collides with a heavy ion, it distorts the ion from its equilibrium position. However, 
since the electron has an energy $\sim E_{F}$, the Fermi energy, it escapes from the vicinity
of the distortion in a time $\sim (\hbar/E_{F})$, while it takes a much longer time $\sim (\hbar/\omega_{D})$,
where $\omega_{D}$ is the Debye energy, for the ion to relax. The distortion polarizes the 
metal, attracting other electrons to it. Crucially, due to the difference in time scales, a second electron
attracted by the distortion experiences only a small repulsion from the initial one, which has escaped 
far away by that time, leading to an effective attraction between the two electrons. 
This reduction in the Coulomb repulsion between the two electrons can be formally expressed 
using various methods, including a renormalization group (RG) approach, leading to the well known Tolmachev-Anderson-Morel (TAM) 
logarithmic reduction \cite{tam-paps} in a clean system in the bulk.

In a diffusive system, the first electron escapes much more 
slowly since it collides frequently with impurities, and may return to the original 
collision area. As a result, the reduction in the Coulomb repulsion is weaker than 
the clean case given by the TAM effect, causing a reduction in $T_{c}$.
In two dimensions, this effect can be formulated in the RG language,
leading to a modified RG equation below the scattering rate ($1/\tau$), 
as shown by Finkel'stein  \cite{finkel-zphys,finkel-physica}.

In a finite size system, the Thouless energy $E_{Th}=(\hbar D/L^{2})$ 
defines another important energy scale. At energies much below $E_{Th}$, 
superconducting systems with dimensionless conductance 
$g=(E_{Th}/\delta) \gg 1$ are described by Richardson's model \cite{richardson}, 
with constant, energy-independent interaction matrix elements 
in the pairing channel. Physically, this expresses the fact that at energies much below $E_{Th}$,
the wavefunctions of all electrons are spread uniformly over the whole 
system, and thus the dynamical component of the electron electron
interaction is no longer present. In the RG language, this leads to
the TAM logarithm again,  resulting in a stronger 
reduction of the Coulomb interaction in this regime, 
similar to a clean system. 
%

In a diffusive, finite size grain, these energy scales form a hierarchy, 
given by $E_{F}>1/\tau>E_{Th}$. Between $E_{F}$ and $1/\tau$, the physics is identical
to that of a bulk clean system, since the electrons are unaware of the disorder and finite size.
Hence, the RG is determined by the TAM equation. Between $1/\tau$ and $E_{Th}$,
the electrons are affected by disorder but not the finite size, and thus follow the Finkel'stein
equation. However, below $E_{Th}$, the finite size effect dominates, and the RG reverts to the TAM equation, 
due to the arguments provided above.

As the grain size $L$ is reduced, $E_{Th}$ is increased,
diminishing the regime where the Finkel'stein effect is relevant, while simultaneously extending
the regime where the TAM equation holds. As a result, in a smaller grain, the Coulomb repulsion
is reduced more strongly since the Finkel'stein regime is smaller, and this should lead to a larger mean field $T_{c}$.

These conclusions are confirmed by our calculations.
We consider isolated grains with size $L \gg t$, where $t$ is
the thickness of the grains, and study the mean field $T_{c}$ as a function of $E_{Th}$ by solving the appropriate
RG equations in the different regimes. Using a specific model for the bare interactions based on the physical arguments
given above, we show that the mean field $T_{c}$
can be increased all the way from the disordered bulk limit, $T^{b}_{c}$, to 
the the clean limit, $T_{c0}$, by simply reducing the size of the 
grain such that $T^{b}_{c}<E_{c}<1/\tau$, where $E_{c}=4 \pi^{2} E_{Th}$. 
We observe an increase of upto $20\%$ in $T_{c}$ when $g \sim {\cal O}(10)$, 
but at the limit of the validity of the theory, when $g \sim {\cal O}(1)$, $T_{c}$ increases by upto $60\%$.

{\it RG in Clean Systems:} Superconductivity is driven by a diverging interaction in the pairing-channel or 
Cooper-channel. In a clean system, the physics
is contained in the repeated scattering of electrons with opposite 
momenta, Matsubara frequencies and spin,  $| \vec{k}\mbox{,} \epsilon_{m} \uparrow \rangle$
and $|\mbox{-}\vec{k}\mbox{,-}\epsilon_{m} \downarrow \rangle$, by the interaction
 \cite{altland}. The screened Coulomb interaction $V_{scr}$ is usually assumed to be local and
instantaneous, and hence the matrix elements for pair scattering between states  $| \vec{k}, \epsilon_{m} \uparrow \rangle$
and $|\mbox{-}\vec{k}\mbox{,-}\epsilon_{m} \downarrow \rangle$ and  $| \vec{k'}\mbox{,} \epsilon_{n} \uparrow \rangle$
and $|\mbox{-}\vec{k'}\mbox{,-}\epsilon_{n} \downarrow \rangle$ is given by a constant $\Gamma^{0}_{mn} = \nu_{0} V_{scr} \approx 1$,
where $\nu_{0}$ is the density of states at the Fermi energy.

The effective phonon-mediated attractive interaction below $\omega_{D}$ is retarded and
hence frequency dependent in general, but is assumed to be a constant for simplicity.
Denoting its value by $\lambda_{a}$, the bare matrix elements
in the clean system are given by

\begin{eqnarray}
 \Gamma^{0}_{mn} &=& 1 \mbox{, } E_{F}>\mbox{max}(\epsilon_{m} \mbox{,} \epsilon_{n} )>\omega_{D} \nonumber \\
                                  &=& (1-\lambda_{a}) \mbox{, }  \epsilon_{m} \mbox{,} \epsilon_{n} < \omega_{D}.
\end{eqnarray}

The full effective interaction $\Gamma_{mn}$ can be found by solving the relevant Bethe Salpeter equation (see discussion on
the disordered case below), or by progressively integrating out thin regions of energy in succession. 
With $\epsilon_{m} {\rm,} \epsilon_{n} \rightarrow \omega$, 
and $\Gamma_{mn} \equiv \Gamma(\omega)$,
both methods lead to the standard TAM RG equation\cite{tam-paps}, given by 

\begin{eqnarray}
 d \Gamma(\omega)/ d l_{\omega} = -\Gamma^{2}(\omega).
\end{eqnarray}

Here $\omega$ is the running energy scale and $l_{\omega} = \mbox{ln}(E_{F}/\omega)$.
Notice that the bare matrix elements do not appear in the equation directly, but only through
the boundary condition $\Gamma(E_{F})=\Gamma_{0}(E_{F}) \approx 1$.
This is easily integrated to give the TAM logarithm reduction \cite{tam-paps}

\begin{eqnarray}
\Gamma(\omega) = \Gamma(E_{F})/(1 + \Gamma(E_{F})\mbox{ln}(E_{F}/\omega)).
\end{eqnarray}

Now the RG proceeds in two steps. First, from $E_{F}$ to $\omega_{D}$, the RG
reduces the effective interaction strongly, according to Eq. (3). 
If the attractive interaction due to the phonons,  $\lambda_{a}$, 
is stronger than the renormalized repulsive interaction $\Gamma(\omega_D)$, then the total
interaction is negative and further renormalization increases in until
it diverges at $\omega = T_{c} \sim \omega_{D} \exp(-1/|\Gamma(\omega_{D})-\lambda_{a}|)$.
In some places in the literature \cite{nagaosa}, $\Gamma[\omega_D]$ is denoted by $\mu^*$.

{\it Disordered:} In disordered systems, one should take into account corrections to the 
$\Gamma_{mn}$ due to disorder. At weak potential disorder, $1/\tau  \ll  \omega_{D}$, 
it is well known that the superconducting $T_{c}$ 
is virtually unchanged \cite{and-tr}. Diagrammatically, this corresponds to incorporating
the effects of disorder and interactions to $\Gamma_{mn}$ {\it separately} \cite{finkel-physica}, i.e., 
where the electron-electron interactions and disorder corrections are factorizable.

However, there is a class of corrections that couple different sections of the matrix elements
 with different indices $m$ and $n$.
 These provide a non-trivial frequency dependence to the resulting matrix elements, 
and the disorder and interaction corrections are no longer factorizable. 
 While these corrections are minor in three dimensions, 
they become important in lower dimensions. 
In two dimensions, the disorder corrected bare Coulomb matrix elements $\Gamma_{mn}$ 
can be explicitly calculated by diagrammatic methods and 
are given by  \cite{finkel-physica,oreg-2d}

\begin{eqnarray}
\Gamma^{0}_{mn} = 1 + ut~ \mbox{ln}(1/(\epsilon_{m}+\epsilon_{n}) \tau) \mbox{, } \epsilon_{n} \mbox{,} \epsilon_{m}<1/\tau.
\end{eqnarray}

Here, $\epsilon_{m}$ and $\epsilon_{n}$ are fermionic Matsubara frequencies, $u\sim0.5$ in two dimensions and  
$t= (1/2 \pi^{2}) (e^{2}/ \hbar) R_{\Box} = 1/(2 \pi^{2} g)$, where $R_{\Box}$ is the sheet resistance in 
two dimensions. The first term is the contribution from the clean bare matrix element, while the frequency 
dependent second term encodes the contribution from disorder corrections.

This form can also be derived by calculating
the pairing channel matrix elements between exact disorder eigenstates $|m \uparrow\rangle$, 
$|m \downarrow\rangle$ and $|n \uparrow\rangle$, 
$|n \downarrow\rangle$. Using semiclassical arguments \cite{gefen},
one can show that

\begin{eqnarray}
 \nu_{0} V_{mn}  &=&  \sum_{q,Dq^{2}<1/\tau} |\langle m | e^{i q r} | n \rangle|^{2} \nonumber \\
                                           &=& (\delta/\pi)\mbox{}\sum_{q,Dq^{2}<1/\tau} D q^{2}/(D^{2} q^{4} + \omega^{2}_{mn}).
\end{eqnarray}

Here, $\omega_{mn}$ is the energy difference between the states $m$ and $n$.
In the continuum limit in two dimensions, the resultant integral yields the logarithm of Eq. (4). 
The full matrix element $\Gamma_{mn}$ is given by the following Bethe Salpeter equation \cite{oreg-2d}:

\begin{eqnarray}
 \Gamma_{mn} = \Gamma^{0}_{mn} - 2 \pi T \sum_{r} \Gamma^{0}_{mr}~  \frac{1}{\epsilon_{r}} ~ \Gamma_{rn}.
\end{eqnarray}

The logarithmic behaviour of $\Gamma^{0}_{mn}$ enables an RG treatment of the system using the 
maximum section approach \cite{noz-maxscn} with the approximation
$\mbox{ln}( (\epsilon_{m}+\epsilon_{n}) \tau) \approx \mbox{ln}(\mbox{max[}\epsilon_{m},\epsilon_{n}\mbox{]} \tau)$.
One then gets a modified RG equation for $\Gamma(\omega)$ given by \cite{finkel-physica,oreg-2d,supp-dir}

\begin{eqnarray}
d \Gamma(\omega)/ d l_{\omega} = u t - \Gamma^{2}(\omega).
\end{eqnarray}

The first term is the nontrivial contribution due to disorder and slows down
the renormalization of the scattering amplitude. In a thermodynamic two dimensional system,
this leads to a suppression of the $T_{c}$ with increasing disorder.

{\it Finite size:} In finite-size systems $E_{Th}$ provides another important energy scale, 
in addition to $E_{F}$, $1/\tau$, $\omega_{D}$ and $T_{c}$, as explained earlier.
It is well known that the statistical properties of the energy eigenstates of such systems for energy scales 
 $\omega  \ll  E_{Th}$ and $g \gg 1$ are described by Random Matrix Theory \cite{rmt}. 
Under these conditions, the system is described
by the so called Universal Hamiltonian \cite{univ-ham} which is determined by three constant coefficients coupling to the 
total density, spin and pairing operators respectively. In situations where only the pairing channel is relevant,
this reduces to the well known Richardson's model \cite{richardson}: 

\begin{eqnarray}
 {\cal H}_{\rm rich} = \sum_{m,\sigma} \epsilon_{m} c^{\dagger}_{m \sigma} c_{m \sigma} +  
\lambda \delta \sum_{m,n} c^{\dagger}_{m \uparrow} c^{\dagger}_{m \downarrow}
 c_{n \downarrow} c_{n \uparrow}
\end{eqnarray}

Here, $m$, $n$ denote the exact eigenstates of the system
and $\lambda$ encodes the strength of the electron-electron interactions. Thus, the matrix elements
become independent of the states they couple, similar to the clean case, and this leads to the TAM RG equations.
To come up with a specific model for the bare matrix elements, we make the crude assumption that 
we can neglect the frequency dependence of the elements below 
$\omega<D q^{2}_{\rm min}= E_{c}$ in Eq. (5). Hence, for $\omega<E_{c}$, 
the bare Coulomb matrix elements assume the constant value 
$\lambda \sim 1 + ut~\mbox{ln} \big( 1/(E_{c} \tau) \big)$.
$E_{c}$ acts as an effective cutoff scale for the bare matrix elements and plays a fundamental
role in the scaling properties of the system.

Hence, in superconducting grains with a relatively large $E_{Th}$ and $L \gg t$,
there are three distinct regimes:

1) $E_{F}>\omega>1/\tau$ : The system is in the ballistic limit, the bare Coulomb matrix elements are
given by $\Gamma^{0}_{mn} = \lambda_{0} = 1$, and the full matrix element $\Gamma$ follows the TAM RG equation, Eq.(4).

2) $1/\tau>\omega>E_{c}$: In this regime, the Coulomb repulsion is affected by disorder, $\Gamma^{0}_{mn}$ is
given by Eq. (4) and the RG is given by the Finkel'stein equation, Eq. (6).

3) $E_{c}>\omega>T_{c}$: In this regime, the matrix elements are effectively constant
and the system can be described by Richardson's model. Using our crude model described earlier,
$\Gamma^{0}_{mn} \sim 1 + ut~\mbox{ln} \big( 1/(E_{c} \tau) \big)$, 
and the system again follows the TAM RG equation.

Of course, for all $(\epsilon_{m}\mbox{,} \epsilon_{n})<\omega_{D}$, the total bare matrix element also includes
the attractive interaction, (-$\lambda_{a}$).

\begin{figure}[b]
\centerline{
\includegraphics[width=6.5cm,height=4.5cm,angle=0]{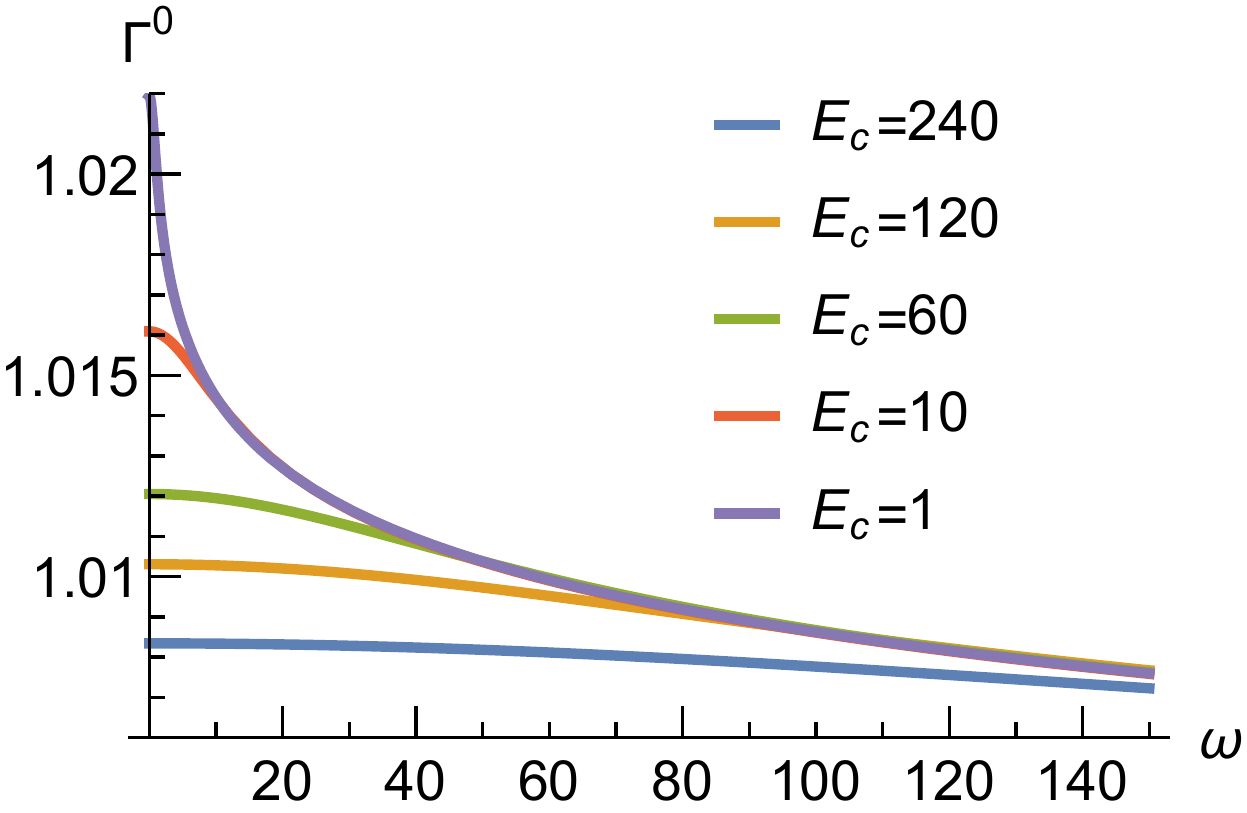}
}
\caption{Colour online: Bare Coulomb matrix elements $\Gamma^{0}$ vs. frequency $\omega$
for different values of $E_{c}=4 \pi^{2} E_{Th}$, where $E_{Th}$ is the Thouless energy, 
at dimensionless conductance $g=10$ and scattering rate $1/\tau=0.1E_{F}$, where $E_{F}$ is the Fermi energy. 
All energy scales are expressed in units of $1K$, which is defined in detail in the text. 
The matrix elements show logarithmic behaviour for $\omega \gtrsim  E_{c}$, consistent
with Eq. (4), but saturate to a constant below this energy scale as it enters the Richardson's regime,
as explained in the text. The corresponding RG equations in these regimes are given
by Eqns. (4) and (2) respectively, as explained in detail in the text.
}
\end{figure} 

These considerations lead to a remarkable conclusion: In the regime $1/\tau>E_{c}>T_{c}$,
 increasing $E_{c}$ by reducing the size of the system diminishes regime 2 and simultaneously extends
regime 3, resulting in a {\it faster} renormalization of the effective interaction. This will lead to an {\it increase}
in the mean field $T_{c}$ of the system, until $E_{c}=1/\tau$, where $T_{c}$ will be equal to the clean limit
value of the material, since the RG would then be determined by the TAM equations throughout the
whole energy range. Thus, within mean field theory,
 one can increase $T_{c}$ {\it all the way} from
the bulk disordered value, $T^{b}_{c}$, (given by the solution of Eq. (7)) to the clean
value $T^{0}_{c}$. 
As mentioned before, this picture ceases to be valid when $\delta \sim \Delta \sim {\cal O}(T_{c})$,
where the superconducting state ceases to exist.

{\it Numerical Results:} We consider isolated grains of materials with transverse dimensions $L$ and thickness $t \ll L$, so that they 
are effectively two dimensional. For direct comparison with real materials, we choose our energy unit such that important
parameters like $E_{F}$ assume values similar to those in real materials expressed in Kelvins ($K$).
We choose reasonable values $E_{F}=30000K$, $1/\tau=3000K=0.1E_{F}$ and
$\omega_{D}=300K=0.01E_{F}$. 
Using typical values for density of states $\nu_{0}$ 
and Fermi velocity $v_{F}$ for free electrons, 
we find a mean free path $l_{e} \approx 2.5$ nm.
Note that for a system with $t=l_{e}$, 
$g=(E_{Th}/\delta) = \hbar D \nu_{0} t \approx 9$.
The maximum value of $E_{c} = (1/\tau)$ implies that the corresponding minimum length 
$L=\sqrt{4 \pi^{2} \hbar D \tau} \approx 9.1$nm. 

\begin{figure}[t]
\centerline{
\includegraphics[width=4.5cm,height=4.0cm,angle=0]{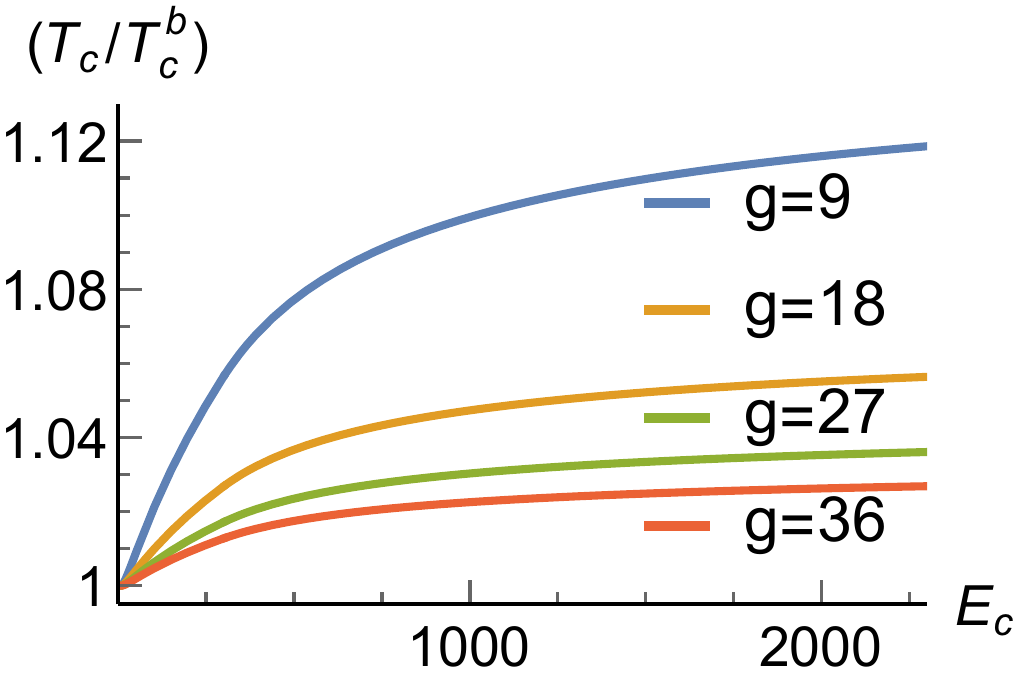}
\hspace{-0.4cm}
\includegraphics[width=4.5cm,height=4.0cm,angle=0]{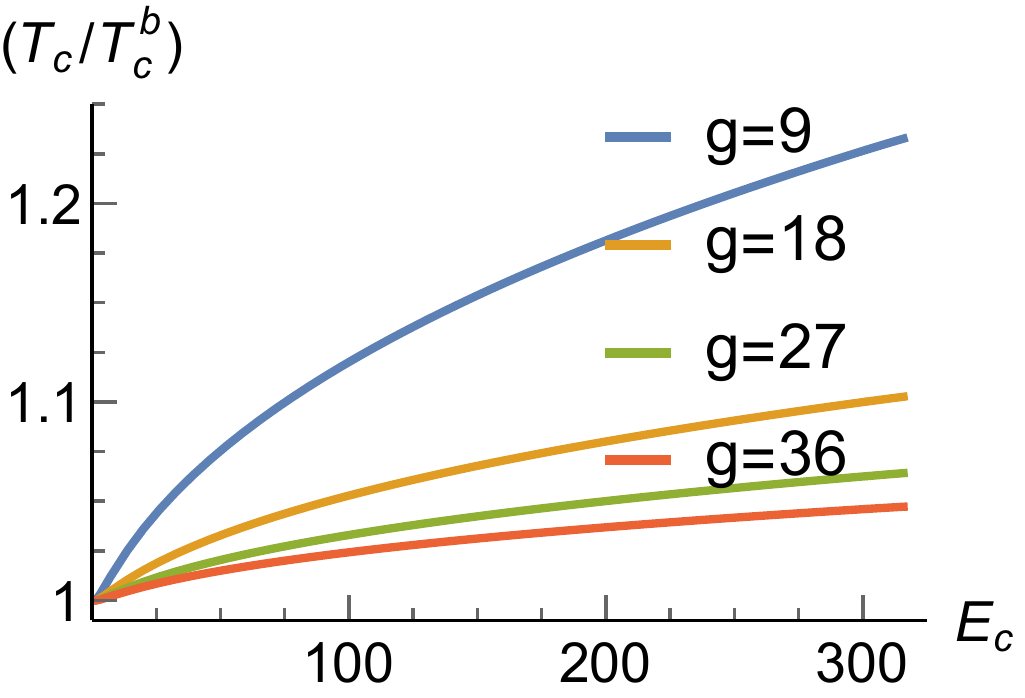}}
\centerline{
\includegraphics[width=4.5cm,height=4.0cm,angle=0]{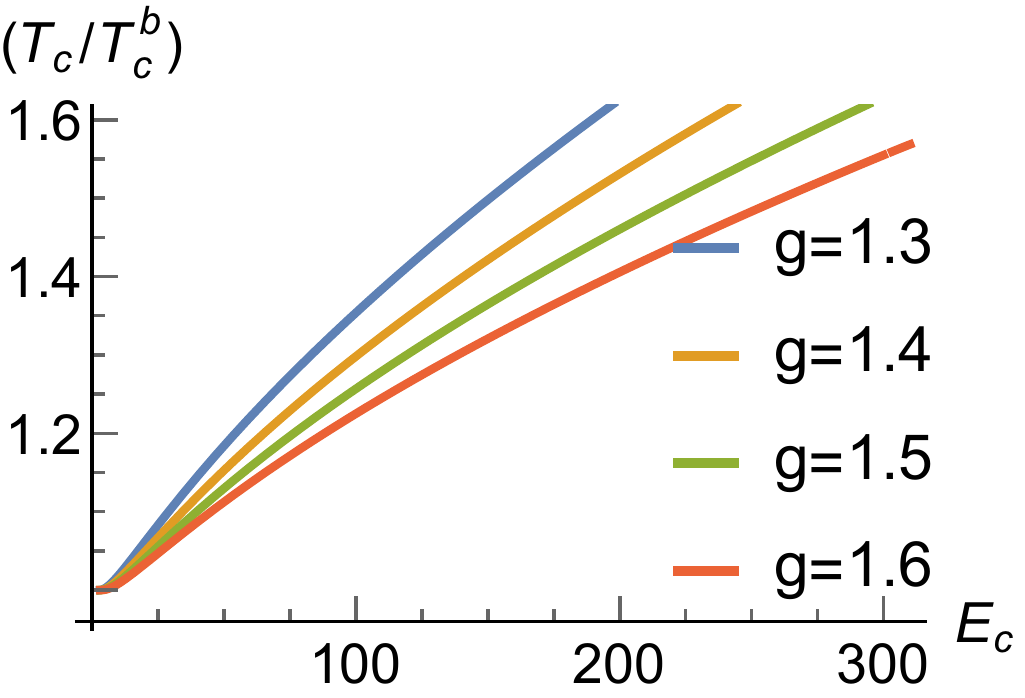}
\hspace{-0.4cm}
\includegraphics[width=4.5cm,height=4.0cm,angle=0]{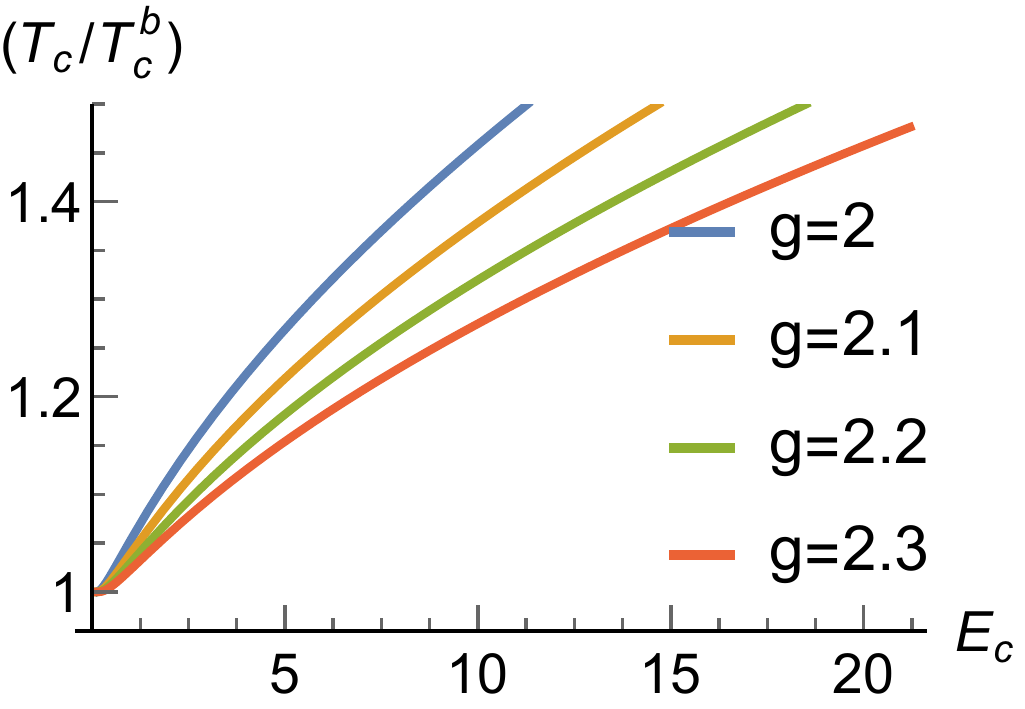}
}
\caption{
Colour online: Normalized $T_{c}$ with increasing $E_{c}=4 \pi^{2} E_{Th}$, where $E_{Th}$
is the Thouless energy, for two different systems with clean limit transition temperature
$T_{c0}=7K$ (left column) and $1K$ (right column) at different values of dimensionless conductance $g$. All
energy scales are expressed in units of $1K$, which is defined in detail in the text. 
The $x$-axis cutoff is determined by 
$\delta \sim T_{c}$, beyond which superconductivity should not survive \cite{and-tr,tc-vanish-paps}.
At large $g \sim {\cal O}(10)$, the small $T_{c}$ system shows a much larger fractional
increase ($20\%$) than the one with large $T_{c}$ ($12\%$). Close to the limit of the validity of the theory, 
at $g \sim {\cal O}(1)$, the enhancement can increase up to $60 \%$.
}
\end{figure} 


First, we analyse the behaviour of the bare matrix elements to gain insight into the RG process.
Fig. 1 shows the bare Coulomb matrix elements as explained in the figure caption in detail. 
The plots demonstrate how these elements, which increase logarithmically
for $\omega> E_{c}$, essentially saturate below this value, validating our crude assumptions in the previous section. 
We have omitted the attractive interaction below $\omega_{D}$ in the plots for clarity. 

Fig. 2 shows our primary result: the normalized transition temperature $T_{c}$ with increasing $E_{c}$
for two cases with $T_{c0} = 7K$ and $1K$ respectively, crudely corresponding to materials with moderately large $T_{c}$ such 
as $Mo\rm{-}Ge$  \cite{mo-ge-ref} and small $T_{c}$ such $Al$.  
We choose the same set of parameters for both except for the attractive interaction parameter $\lambda_{a}$, which is adjusted 
to yield the respective values of $T_{c}$.
We find that at large $g \sim 10$, 
the system with the larger $T_{c}$ shows an increase of $12\%$, while the 
one with the smaller $T_{c}$ shows a much larger increase of $20\%$. 
Hence, grains with small $T_{c}$ show a much larger fractional increase with reducing size at large $g$,
which is a correlation borne out by experiments.
Furthermore, by pushing the theory close to its limit of validity $g \sim {\cal O}(1)$,
we get an enhancement close to $60\%$ in both cases.
We reiterate that this remarkable conclusion follows simply from examining the RG flow of the system
with various values of $E_{c}$, with no reference whatsoever to specific details of the 
material parameters and its geometry. 
%

{\it Experimental verification and Discussion:} 
Our theory concerns isolated grains, studied in some experiments \cite{garciaab}, whereas other experimental 
samples consist of an array of such grains coupled by an effective Josephson coupling. This provides a new energy scale
in these systems whose collective properties, including transport, 
may be very different from those of individual grains. 
Hence, to verify our predictions experimentally, one must focus on isolated or weakly 
coupled grains, and measurements sensitive to the 
single particle indicators of the superconducting state of individual grains, such as the local gap.
Examples of the above are the specific heat 
capacity, which shows a peak at the superconducting transition,
and scanning tunneling measurements, which  can 
measure the local density of states, and thus the local gap, directly.
Hence, we propose that our predictions should be verifiable from measurements
of specific heat capacity and tunneling spectra to track the transition in individual grains.
 
As discussed earlier, we have neglected various other finite size 
effects discussed in the literature that may lead to an 
enhancement of the $T_{c}$ in relatively clean grains. 
Since our mechanism is completely different from these, our effect will be 
superposed on all these in real disordered grains. This seems to be relevant in
granular $Al$ films, for example, where the $T_{c}$ can substantially exceed 
$T_{c0}$ \cite{deutscher2}, but here our mechanism may contribute a part of 
the total increase.
Furthermore, we have neglected the progressive broadening of the superconducting
transition with increasing $\delta / T_{c}$ (leading to its eventual disappearance in the 
$\delta / T_{c} \rightarrow 1$ limit \cite{and-tr,tc-vanish-paps}), analysed in detail in Ref.\cite{muhl}.
These results seem to indicate that the broadening is not accompanied by an appreciable shift in the transition
in small grains.
We have also not discussed the effects of fluctuations in the grain sizes in experiments
on distributions of grains \cite{garciac}.

%
In conclusion, by considering the RG equations in different energy regimes, we have demonstrated a 
universal mechanism for increasing the superconducting $T_{c}$ 
in isolated disordered grains with reducing size, from the bulk value $T^{b}_{c}$ to the clean 
limit $T_{c0}$ in the regime $T_{c}<E_{c} < 1/\tau$. 
This prediction can be tested experimentally by 
measuring properties sensitive to the local single particle gap
such as the specific heat capacity. 

We acknowledge enlightening discussions with A. M. Finkel'stein, A. D. Mirlin, A. Kamenev and Z. Ovadyahu,
financial support by the Israel Science Foundation and the Weizmann Institute.


\clearpage

\include{coul-ren-prl-supp}

\end{document}

%% file: coul-ren-prl-supp.tex
\bibliographystyle{apsrev4-1}

\makeatletter
\renewcommand{\thefigure}{S\arabic{figure}}
\renewcommand{\figurename}{Figure}

\begin{widetext}
\begin{center}
\large{\bf Supplementary material for\\
 ``Reduction of electron repulsion and enhancement 
of $T_{c}$ in small diffusive superconducting grains''}
\end{center}

\begin{center}
\large{Sabyasachi Tarat, Yuval Oreg and Yoseph Imry} 
\end{center}

\section{Derivation and solution of RG equations}

The Bethe Salpeter equation for the Cooper channel is given by

\begin{eqnarray}
 \Gamma_{mn} = \Gamma^{0}_{mn} - 2 \pi T \sum_{r} \Gamma^{0}_{mr}~  \frac{1}{\epsilon_{r}} ~ \Gamma_{rn}.
\end{eqnarray}

Fig. S1 below shows schematically the structure of the equation and explains the different components.

\vspace{1cm}
\begin{figure}[h]
\centerline{
\includegraphics[width=14cm,height=2.0cm,angle=0]{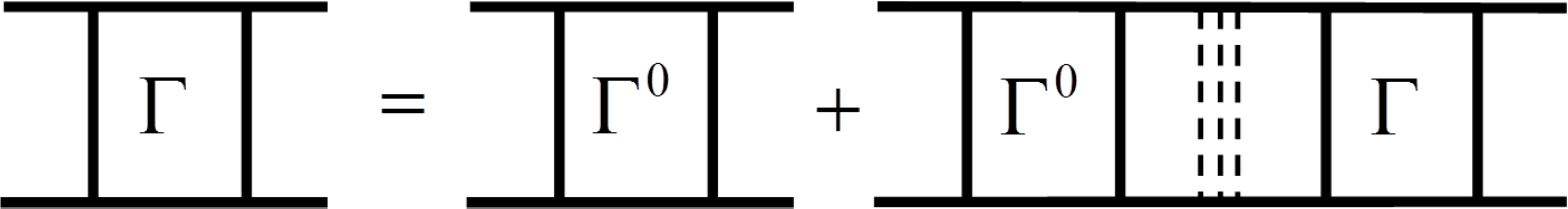}
}
\caption{The Bethe Salpeter equation for the full
matrix elements $\Gamma_{mn}$ in the disordered system. The block $\Gamma^{0}$ is the 
bare matrix element given by $\Gamma^{0}_{mn} = 1 + (u / 2 \pi^{2} g) {\rm ln}(1/(\epsilon_{m}+\epsilon_{n}) \tau)$.
As explained in the main text, the first term is the clean bare Coulomb matrix element (alongwith the attractive interaction below $\omega_{D}$) while the second term includes the disorder 
corrections. The relevant diagrams that contribute to the second term are provided in Ref.\cite{finkel-physica} and Ref.\cite{oreg-2d}. Detailed calculations for 
similar diagrams yielding the logarithmic dependence may be found in Ref.\cite{maekawa-diag}. 
The dashed lines are impurity corrections and lead to a contribution $\sim (\pi \nu_{0} / |\epsilon_{r}|)$. Plugging these
values into the diagram, one gets Eq.(1), where the factor of $2$ comes from the fact that the sum is only over positive values of $\epsilon_{r}$.
}
\end{figure}

To solve this equation in the disordered case, we focus on an energy scale $\epsilon \sim {\cal O}(T)$ and look at $\Gamma(\epsilon, \epsilon) \equiv \Gamma(\epsilon)$.
Converting the sum in Eq.(1) to an integral, and noting that the Matsubara sum implies a lower cutoff $\sim \epsilon$ in the integral, we have 

\begin{eqnarray}
\Gamma(\epsilon) = \Gamma^{0}(\epsilon) - \int_{\epsilon} d \omega ~\Gamma^{0}(\epsilon, \omega) ~ \frac{1}{\omega}~ \Gamma(\omega, \epsilon)
\end{eqnarray}

Now, we use the fact that $\Gamma^{0}(\epsilon, \omega) \propto {\rm ln}(1/(\epsilon + \omega) \tau) \approx  {\rm ln}(\mbox{max[}\epsilon,\omega \mbox{]} \tau) = {\rm ln}(1/(\omega) \tau) \equiv \Gamma^{0}(\omega)$.
It is obvious that the full element $\Gamma(\epsilon, \omega) \approx \Gamma(\omega)$ as well. As explained in the appendix of Ref.\cite{noz-maxscn}, the next step is to expand the right hand side, and look
at the individual diagrams, which consist of individual $\Gamma^{0}$ blocks joined by the impurity sections. In a given diagram, we identify the impurity section with the largest energy, say $\omega$.
Now, as long as all other energies are less than $\omega$, any number of $\Gamma^{0}$ blocks and impurity sections may be inserted both to the left and right of this section to create a valid term
in the series. This implies that after summing them all up, we get back the full $\Gamma(\omega)$ both on the left and the right (with no overcounting in this case), and Eq.(2) can be rewritten as

\begin{eqnarray}
\Gamma(\epsilon) = \Gamma^{0}(\epsilon) - \int_{\epsilon} d \omega ~\frac{\Gamma^{2}(\omega)}{\omega}
\end{eqnarray}

Now, taking the derivative w.r.t ${\rm ln}(1/(\epsilon \tau))$, we get 

\begin{eqnarray}
 d \Gamma(\epsilon)/ d l_{\epsilon} = u/(2 \pi^{2} g) - \Gamma^{2}(\epsilon) {\rm ,}
\end{eqnarray}

which is the required RG equation \cite{finkel-physica,oreg-2d}.

The solution of this equation is given by 

\begin{eqnarray}
\Gamma(\omega) = \sqrt{ut} \Bigg\{ \frac{1 + \Big(  \frac{1 - \sqrt{ut/ \Gamma_{0}}}{1 + \sqrt{ut/ \Gamma_{0}}} \Big) \Big( \frac{\omega}{\omega_{0}}   \Big)^{2 \sqrt{ut}}   }{1 - \Big(  \frac{1 - \sqrt{ut/ \Gamma_{0}}}{1 + \sqrt{ut/ \Gamma_{0}}} \Big) \Big( \frac{\omega}{\omega_{0}}   \Big)^{2 \sqrt{ut}}} \Bigg\}
\end{eqnarray}

where $\omega_{0}$ is the high energy cutoff where the matrix element is $\Gamma_{0}$, and $t = 1/(2 \pi^{2} g)$. The value of $\omega$ where the denominator vanishes gives the $T_{c}$ 
in the Finkel'stein theory.

\end{widetext}

%% file: coul-ren-prl.bbl
\begin{thebibliography}{99}
\bibitem{fn1} In this context, we mention that 
starting from the classic experimental work of Buckel and Hilsch \cite{bh-paps}, 
it has been well known (though poorly understood)
that disorder can sometimes raise $T_{c}$ in quench condensed amorphous materials.
Recent work \cite{burmistrov} has shown that in bulk systems where the Coulomb
repulsion is screened strongly  (e.g. by an external metallic plate close to a 2D disordered system), 
disorder can increase $T_{c}$ due to the multifractal structure of the 
electronic wavefunctions and the effect of the direct and exchange channels on the Cooper channel.
We do not address these issues here.
Our theoretical considerations show that in dirty systems with Coulomb
interactions that are screened only by the system itself, 
where disorder decreases $T_{c}$, finite size can actually negate that and raise $T_{c}$
back.
\end{thebibliography}
